\documentclass[prodmode]{acmsmall} 

\acmVolume{0}
\acmNumber{0}
\acmArticle{0}
\acmYear{2015}
\acmMonth{10}
\usepackage{glossaries}

\doi{0000001.0000001}

\usepackage{color}
\usepackage{booktabs}
\usepackage{amssymb}

\usepackage{pgfplots,url}

\definecolor{mycol1}{HTML}{0072BD}
\definecolor{mycol2}{HTML}{D95319}
\definecolor{mycol3}{HTML}{EDB120}
\definecolor{mycol4}{HTML}{7E2F8E}
\definecolor{mycol5}{HTML}{77AC30}
\definecolor{mycol6}{HTML}{4DBEEE}
\definecolor{mycol7}{HTML}{A2142F}

\begin{document}

\markboth{I. \v{Z}liobait\.e}{Discrimination measures}

\title{A survey on measuring indirect discrimination in machine learning} 
\author{INDR\.E \v{Z}LIOBAIT\.E
\affil{Aalto University and Helsinki Institute for Information Technology HIIT}
}

\begin{abstract}

Nowadays, many decisions are made using predictive models built on historical data. 
Predictive models may systematically discriminate groups of people even if the computing process is fair and well-intentioned.
Discrimination-aware data mining studies how to make predictive models free from discrimination, when historical data, on which they are built, may be biased, incomplete, or even contain past discriminatory decisions. 
Discrimination refers to disadvantageous treatment of a person based on belonging to a category rather than on individual merit. 
In this survey we review and organize various discrimination measures that have been used for measuring discrimination in data, as well as in evaluating performance of discrimination-aware predictive models. 
We also discuss related measures from other disciplines, which have not been used for measuring discrimination, but potentially could be suitable for this purpose. We computationally analyze properties of selected measures. 
We also review and discuss measuring procedures, and present recommendations for practitioners. 
The primary target audience is data mining, machine learning, pattern recognition, statistical modeling researchers developing new methods for non-discriminatory predictive modeling.
In addition, practitioners and policy makers would use the survey for diagnosing potential discrimination by predictive models.
\end{abstract}


\terms{fairness in machine learning, predictive modeling, non-discrimination, discrimination-aware data mining}



%

\maketitle

\newpage

\tableofcontents

\newpage
\section{Introduction}

Nowadays, many decisions are made using predictive models built on historical data, for instance, personalized pricing and recommendations, credit scoring, automated CV screening of job applicants, profiling of potential suspects by the police, and many more. Penetration of machine learning technologies, and decisions informed by big data has raised public awareness that automated decision making may lead to discrimination \cite{Obama_report,nytimes,guardian}.
Predictive models may discriminate people, even if the computing process is fair and well-intentioned \cite{Barocas16,Citron14,Calders13why}.
This is because most machine learning methods are based upon assumptions that the historical data is correct, and represents the population well, which is often far from reality.

Discrimination-aware machine learning and data mining is an emerging discipline, which studies how to prevent discrimination in predictive modeling. It is assumed that non-discrimination regulations, such as which characteristics, or which groups of people are considered as protected, are externally defined by national and international legislation. 
The goal is to mathematically formulate non-discrimination constraints, and develop machine learning algorithms that would be able to take into account those constraints, and still be as accurate as possible.

In the last few years researchers have developed a number of discrimination-aware machine learning algorithms, using a variety of performance measures. 
Nevertheless, there is a lack of consensus how to define fairness of predictive models, and how to measure the performance in terms of discrimination. 
Quite often research papers propose a new way to quantify discrimination, and a new algorithm that would optimize that measure. The variety of approaches to evaluation makes it difficult to compare the results and assess the progress in the discipline, and even more importantly, it makes it difficult to recommend computational strategies for practitioners and policy makers.

The goal of this survey is to present a unifying view towards discrimination measures in machine learning, and understand the implications of choosing to optimize one or another measure, because measuring is central in formulating optimization criteria for algorithmic discrimination discovery and prevention.
Hence, it is important to have a structured survey at an early stage of development of this research field, in order to present task settings in a systematic way for follow up research, and to enable systematic comparison of approaches. 
Thus, we review and categorize measures that have been used in machine learning and data mining, and also discuss existing measures from other fields, such as feature selection, which in principle could be used for measuring discrimination. 

There are several related surveys that can be viewed as complementary to this survey. 
A recent review \cite{Romei14} presents a multi-disciplinary context for discrimination-aware data mining. This survey contains a brief overview of discrimination measures with does not go into analysis and comparison of the measures, since the focus is on approaches to solutions across different disciplines (law, economics, statistics, computer science). 
Another recent review \cite{Barocas16} discusses legal aspects of potential discrimination by machine learning, mainly focusing on American anti-discrimination law.
A matured handbook on measuring racial discrimination \cite{Blank04} focuses on surveying and collecting evidence for discrimination discovery. 
The book is not considering discrimination by algorithms, only by human decision makers.


The remainder of the article is organized as follows. 
Section \ref{sec:legislation} presents legal context, terminology, and 
provides an overview of research in developing non-discriminatory predictive modeling approaches. Our intention is to keep this section brief. An interested reader is referred to focused surveys \cite{Romei14,Barocas16} for more information. 
Section \ref{sec:measures} reviews and organizes discrimination measures used in discrimination-aware machine learning and data mining, as well as potentially useful measures from other fields. 
Section \ref{sec:analysis} analyzes and compares a set of most popular measures, and discusses implications of using one or the other. 
Finally, Section \ref{sec:conclusion} presents recommendations for researchers, and concludes the survey.

\section{Background}
\label{sec:legislation}

\subsection{Discrimination and law}

Discrimination translates from latin as \emph{a distinguishing}. 
While distinguishing is not wrong as such, discrimination has a negative connotation referring to adversary treatment of people based on belonging to some group rather than individual merits. 
Public attention to discrimination prevention has been increasing in the last few years. 
National and international anti-discrimination legislation are extending the scope of protection against discrimination, and expanding discrimination grounds. 

Adversary discrimination is undesired from the perspective of basic human rights, and in many areas of life non-discrimination is enforced by international and national legislation, 
to allow all individuals an equal prospect to access opportunities available in a society \cite{FRA_law}. Enforcing non-discrimination is not only for benefiting individuals. 
Considering individual merits rather than group characteristics is expected to benefit decision makers as well leading to more more informed, and thus likely more accurate decisions. 

Discrimination can be characterized by three main concepts: (1) what actions (2) in which situations (3) towards whom are considered discriminatory. 
Actions are forms of discrimination, situations are areas of discrimination, and grounds of discrimination describe characteristics of towards whom discrimination may occur. 

For example, the main grounds for discrimination defined in European Council directives  \cite{EC_law} (2000/43/EC, 2000/78/EC) are: race and ethnic origin, disability, age, religion or belief, sexual orientation, gender, nationality. Multiple discrimination occurs when a person is discriminated on a combination of several grounds.
The main areas of discrimination are: access to employment, access to education, employment and working conditions, social protection, access to supply of goods and services.

Discriminatory actions may take different forms, the two main of which are known as direct discrimination and indirect discrimination.
A direct discrimination occurs when a person is treated less favorably than another is, has been or would be treated in a comparable situation on protected grounds. For example, property owners are not renting to a minority racial tenant. 
An indirect discrimination (also known as structural discrimination) occurs where an apparently neutral provision, criterion or practice would put persons of a protected ground at a particular disadvantage compared with other persons. For example, a requirement to produce an ID in a form of driver's license for entering a club may discriminate visually impaired people, who cannot have a driver's license.
A related term \emph{statistical discrimination} \cite{Arrow73} is often used in economic modelling. It refers to inequality between demographic groups occurring even when economic agents are rational and non-prejudiced. 

Indirect discrimination applies to machine learning and data mining, since algorithms produce decision rules or decision models. 
While human decision makers may make biased decisions on case by case basis, rules produced by algorithms are applied consistently, and may discriminate more systematically and at a larger scale. 
Discrimination due to algorithms is sometimes referred to as \emph{digital discrimination} (e.g. \cite{Wihbey15}) . 


General population, and even many data scientists may think that algorithms are based on data, and, therefore, models produced by algorithms are always objective.
However, models are as objective as the data on which they are applied, and as long as the assumptions behind the models perfectly match the reality. In practice, this is rarely the case. 
Historical data may be biased, incomplete, or record past discriminatory decisions that can easily be transferred to predictive models, and reinforced in new decision making \cite{Calders13why}. 
Lately, awareness of policy makers and public attention to potential discrimination has been increasing  \cite{Obama_report,nytimes,guardian}, but there is a long way ahead before we can fully understand how such discrimination happens and how to prevent it.

\subsection{Discrimination-aware machine learning and data mining}

Non-discriminatory machine learning and data mining, a discipline at an intersection of computer science, law and social sciences, focuses on two main research directions: \emph{discrimination discovery}, and \emph{discrimination prevention}. 
Discrimination discovery aims at finding discriminatory patterns in data using data mining methods. 
Data mining approach for discrimination discovery typically mines association and classification rules from the data, and then assesses those rules in terms of potential discrimination \cite{Ruggieri10,Romei12,Hajian13,Pedreschi12,Luong11,Mancuhan14}.
A more traditional statistical approach to discrimination discovery typically fits a regression model to the data including the protected features (such as race, gender), and then analyzes the magnitude and statistical significance of the regression coefficients at the protected attributes (e.g. \cite{Edelman14}). If those coefficients appear to be significant, then discrimination is flagged.  

Discrimination prevention develops machine learning algorithms that would produce predictive models, ensuring that those models are free from discrimination,
while, standard predictive models, induced by machine learning and data mining algorithms, may discriminate groups of people due to training data being biased, incomplete, or recording past discriminatory decisions. The goal is to have a model (decision rules) that would obey non-discrimination constraints, typically the constraints directly relate to the selected discrimination measure. 
Solutions for discrimination prevention in predictive models fall into three categories: data preprocessing, model postprocessing, and model regularization.
Data preprocessing modifies the historical data such that the data no longer contains discrimination, and then uses regular machine learning algorithms for model induction. 
Data preprocessing may modify the target variable \cite{Kamiran09,Mancuhan14,KamiranC13}, or modify input data \cite{Feldman15,Zemel13}.
Model postprocessing produces a regular model and then modifies it (e.g. by changing the labels of some leaves in a decision tree) \cite{Kamiran10,Calders10}.
Model regularisation adds optimization constraints in the model learning phase (e.g. by modifying the splitting criteria in decision tree learning) \cite{Kamiran10,Calders13,Kamishima12}.
An interested reader is invited to consult an edited book \cite{Toon_book}, a special issue in a journal \cite{Salvatore_journal}, and proceedings of three workshops in discrimination-aware data mining and machine learning \cite{DAPDM,FATML14,FATML15} for more details.


Defining coherent discrimination measures is central for both lines of research: discrimination discovery and discrimination prevention.
Discrimination discovery needs a measure in order to judge whether there is discrimination in data. Discrimination prevention needs a measure as an optimization criteria in order to sanitize predictive models. Hence, our main focus in this survey is to review discrimination measures, and analyze their properties, and understand implications of using one or another measure. 


\section{Machine learning settings, definitions and scenarios}
\label{sec:tasks}

 \subsection{Definition of fairness for machine learning}
 \label{sec:def}


In the context of machine learning non-discrimination can be defined as follows: 
\textbf{(1) people that are similar in terms non-protected characteristics should receive similar predictions, and 
(2) differences in predictions across groups of people can only be as large as justified by non-protected characteristics.}

The first condition relates to direct discrimination, and can be illustrated by so called \emph{twin test}: if gender is the protected attribute and we have two identical twins that share all characteristics, but gender, they should receive identical predictions. 
The first part is necessary but not sufficient condition to make sure that there is no discrimination in decision making. 

The second condition ensures that there is no indirect discrimination, also referred to as \emph{redlining}. For example, banks used to deny loans for residents of selected neighborhoods. 
Even though race was not formally used as a decision criterion, it appeared that the excluded neighborhoods had much higher population of non-white people than average. 
Even though people from the same neighborhood ("twins") are treated the same way no matter what the race is, artificial lowering of positive decision rates in the non-white-dominated neighborhoods would harm the non-white population more than white. Therefore, different decision rates across neighborhoods can only be as large as justified by non-protected characteristics, and this is what the second part of the definition controls. 

More formally, let $X$ be a set of variables describing non-protected characteristics of a person,  $S$ be a set of variables describing the protected characteristics, and $\hat{y}$ be the model output. A predictive model can be considered fair if: 
(1) the expected value for model output does not depend on the protected characteristics $E(\hat{y}|X,S) = E(\hat{y}|X)$ for all $X$ and $S$, that is, there is no direct discrimination; and 
(2) if non-protected characteristics and protected characteristics are not independent, then the expected value for model output dependence on those non-protected characteristics should be justified, that is if $E(X|S) \neq E(X)$, then $E(\hat{y}|X) = e^\star(\hat{y}|X)$, where $e^\star$ is a constraint. 

Finding and justifying $e^\star$ is non-trivial and very challenging, and that is where a lot of ongoing effort in discrimination-aware machine learning concentrate. 


\subsection{Machine learning task settings}

Machine learning settings for decision support, where discrimination may potentially occur, can take many different forms. 
The variable that is to be predicted -- target variable -- may be binary, ordinal, or numeric, corresponding to binary classification, multiclass classification or regression tasks. 
As an example of a binary classification task in the banking domain could be deciding whether to accept or decline loan application of a person. 
Multiclass classification task could be to determine to which customer benefit program a person should be assigned (e.g. "golden clients", "silver clients", "bronze clients"). 
Regression task could be to determine the interest rate for a particular loan for a particular person. 

Discrimination can occur only when target variable is polar. That is, each task setting some outcomes should be considered superior to others. For example, getting a loan is better than not getting a loan, or the "golden client" package is better than the "silver", and "silver" is better than "bronze", or assigned interest rate $3\%$ is better than $5\%$. If the target variable is not polar, there is no discrimination, because no treatment is superior or inferior to other treatment. 


The protected characteristic, in machine learning settings referred to as the protected variable or sensitive attribute, may as well be binary, categorical or numeric, and it does not need to be polar. 
For example, gender can be encoded with a binary protected variable,  ethnicity can be encoded with a categorical variable, and age can be encoded with  a numerical variable. 
In principle, any combination one or more personal characteristics may be required to be protected. 
Discrimination on more than one ground is known as \emph {multiple discrimination}, and it may be required to ensure prevention of multiple discrimination  in predictive models. 
Thus, ideally, machine learning methods and discrimination measures should be able to handle any type or a combination of protected variables.
For instance, the authorities may want to enforce non-discrimination with respect to ethnicity in determining interest rate, 
or non discrimination with respect to gender \emph{and} age in deciding whether to accept loan applications. 
In discrimination prevention it is assumed that the protected ground is externally given, for example, by law. 

 \subsection{Principles for making machine learning non-discriminatory}

\begin{figure}
\centering
\includegraphics[width=0.6\textwidth]{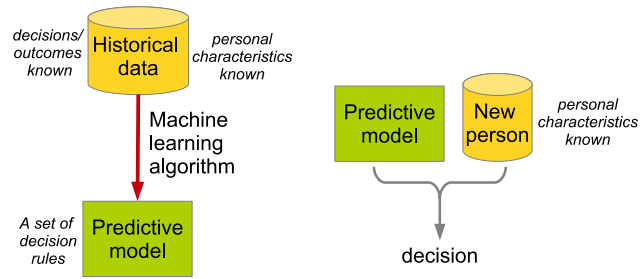}
\caption{A typical machine learning setting.}
\label{fig:ml}
\end{figure}

A typical machine learning process is illustrated in Figure \ref{fig:ml}. 
A machine learning algorithm is a procedure used for producing a predictive model from historical data. 
A model is the resulting decision rule (or a collection of rules). 
The resulting model is used for decision making for new incoming data. 
The model would take personal characteristics as inputs (for example, income, credit history, employment status), and output a prediction (for example, credit risk level). 

Algorithms themselves do not discriminate, because they are not used for decision making. Models (decision rules) that are used for decision making may potentially discriminate people with respect to certain characteristics. 
Algorithms, on the other hand, may be discrimination-aware by employing specific procedures during model construction to enforce non-discriminatory constraints into the models. 
Hence, one of the main goals of discrimination-aware machine learning and data mining is to develop discrimination-aware algorithms, that would guarantee that non-discriminatory models are produced. 
 
There is an ongoing debate in the discrimination-aware data mining and machine learning community whether models should or should not use protected characteristics as inputs. For example, a credit risk assessment model may use gender as input, or may leave the gender variable out. 
Our position on the matter is as follows. Using the protected characteristic as model input may help to ensure that there is no indirect discrimination (for example, as demonstrated in the experimental section of \cite{}). However, if a model uses the protected characteristic as input, the model is not treating two persons that share identical characteristics except for the protected characteristic the same way, a direct discrimination would be propagated. 
Therefore, such a model would be discriminatory discriminatory due to violation of condition \#1 in the definition in Section \ref{lab:df}.
Hence, the model should not use the protected characteristic for decision making.

However, we see no problem in using the protected characteristic in the model learning process, which often may help to enforce non-discrimination constraints. Thus, machine learning algorithms can use the protected characteristic in the learning phase, as long as the resulting predictive model does not require the protected characteristic when used for decision making.

Ensuring that there is no indirect discrimination is much more tricky. In order  to verify to what extent non-discriminatory constraints are obeyed, 
and enforce fair allocation of predictions across groups of people, machine learning algorithms must have access to  the protected characteristics in the historical data. We argue that if protected information (e.g. gender or race) is not available during the model learning building process, the learning algorithm cannot be discrimination-aware, because it cannot actively control non-discrimination. 
The resulting models produces without access to sensitive information may be discriminatory, may be not, but that is by chance rather than discrimination-awareness  property of the algorithm.
 
Non-discrimination can potentially be measured on data (historical data), on predictions made by models, or on models themselves. 
Different task settings and application goals may require different measurement techniques. 
In order to select appropriate measures, which also typically serve as optimisation constraints in the non-discriminatory model learning process, it is important to understand underlying assumptions and basic principles behind different discrimination measures. 
The next section presents a categorized survey of measures used in the discrimination-aware data mining and machine learning literature, and discusses other existing measures that could in principle be used for measuring fairness of algorithms. 
The goal is to present arguments for selecting relevant measures for different learning settings. 



%
%

%
%
%
%
%
%
%
%

%
%

\section{Discrimination measures}
\label{sec:measures}

Discrimination measures can be categorized into (1) statistical tests, (2) absolute measures, (3) conditional measures, and (4) structural measures.
We survey measures in this order due to historical reasons, which is more or less how they came into use. First statistics tests were used which would answer yes or no, then absolute measures came into play that allow quantifying the extent of discrimination, then conditional measures appeared that take into account possible legitimate explanations of differences between different groups of people. Statistical tests, absolute measures and conditional measures are designed for indicating indirect discrimination.
Structural measures have been introduced mainly in accord to mining classification rules, aiming at discovering direct discrimination, but in principle they can also address indirect discrimination.
All these types are not intended as alternatives, but rather reflect different aspects of the problem, as summarized in Table \ref{tab:measuretypes}. 
\begin{table}
\tbl{Discrimination measure types\label{tab:measuretypes}}{%
\begin{tabular}{lll}
\hline
Measures & Indicate what? & Type of discrimination \\
\hline
Statistical tests & presence/absence of discrimination & indirect \\
Absolute measures & magnitude of discrimination & indirect \\
Conditional measures & magnitude of discrimination & indirect \\
Structural measures & spread of discrimination & direct or indirect \\
\hline
\end{tabular}
}
\end{table}

Statistical tests indicate presence or absence of discrimination at a dataset level, they do not measure the magnitude of discrimination, neither the spread of discrimination within the dataset.
Absolute measures capture the magnitude of discrimination over a dataset taking into account the protected characteristic, and the prediction decision; no other characteristics of individuals are considered. 
It is assumed that all individuals are alike, and there should be no differences in decisions for the protected and the general group of people, disregarding any possible explanation.
Absolute measures generally are not for using stand alone on a dataset, but rather provide core principles for conditional measures, or statistical tests. 
Conditional measures capture the magnitude of discrimination, which cannot be explained by any non-protected characteristics of individuals. Statistical tests, absolute and conditional measures are designed to capture indirect discrimination at a dataset level. 
Structural measures do not measure the magnitude of discrimination, but the spread of discrimination, that is, a share of people in the dataset that are affected by direct discrimination. 

Our survey of measures will use mathematical notation as summarized in Table \ref{tab:notation}.
For simplicity we will use the following short probability notation:
$p(s=1)$ will be encoded as $p(s^1)$, and $p(y=+)$ will be encoded as $p(y^{+})$. 
Let $s^1$ denote the protected community, and $y^+$ denote the desired decision (e.g. positive decision to grant a loan).
Upper indices will denote values, lower indices will denote counters of variables.

\begin{table}
\tbl{Solutions\label{tab:notation}}{%
\begin{tabular}{ll}
\hline
Symbol & Explanation \\
\hline
$y$ & target variable, $y_i$ denotes the $i^{th}$ observation \\
$y^i$ & a value of a binary target variable, $y \in \{y^+,y^-\}$\\
$s$ & protected variable \\
$s^i$ & a value of a discreet/binary protected variable, $s \in \{s^1,\ldots,s^m\}$ \\
& typically index $1$ denotes a protected group, e.g. $s^1$ - black, $s^0$ - white race\\
$X$ & a set of input variables (predictors), $X = \{x^{(1)},\ldots,x^{(l)}\}$ \\
$z$ & explanatory variable or stratum \\
$z^i$ & a value of explanatory variable $z \in \{z^1,\ldots,z^k\}$ \\
$N$ & number of individuals in the dataset \\
$n_i$ & number of individuals in group $s^i$ \\
\hline
\end{tabular}
}
\end{table}

\subsection{Statistical tests}

Statistical tests are the earliest measures for indirect discrimination discovery in data. 
Statistical tests are formal procedures to accept or reject statistical hypotheses, which check how likely the result is to have occurred by chance. In discrimination analysis typically the null hypothesis, or the default position, is that there is no difference between the treatment of the general group and the protected group. The test checks, how likely the observed difference between groups has occurred by chance. If chance is unlikely then the null hypothesis is rejected and discrimination is declared. 

Two limitations of statistical tests need to be kept in mind when using them for measuring discrimination. 
\begin{enumerate}
\item Statistical significance does not mean practical significance; statistical tests do not show the magnitude of the the differences between the groups, which can be huge, or can be minor.
\item If the null hypothesis is rejected then discrimination is present, but if null hypothesis cannot be rejected, this does not prove that there is no discrimination. It maybe that the data sample is too small to declare discrimination. 
\end{enumerate}

Standard statistical tests are typically applied for measuring discrimination. The same tests are used in clinical trials, marketing, and scientific research. 

\subsubsection{Regression slope test}

The test fits an ordinary least squares (OLS) regression to the data including the protected variable, and tests whether the regression coefficient of the protected variable is significantly different from zero. A basic version for discrimination discovery considers only the protected characteristic $s$ and the target variable $y$ \cite{Yinger86}. In principle $s$ and $y$ can be binary or numeric, but typically in discrimination testing $s$ is binary. The regression may include only the protected variable $s$ as a predictor, but it may also include variables from $X$ that may explain some of the observed difference in decisions.

The test statistic is $t = b/\sigma$, where $b$ is the estimated regression coefficient of $s$, and $\sigma$ is the standard error, computed as 
$\sigma = \frac{\sqrt{\sum_{i=1}^n (y_i - f(y_i))^2}}{\sqrt{(n-2)}\sqrt{\sum_{i=1}^n (s_i - \bar{s})^2}}$, where $n$ is the number of observations, $f(.)$ is the regression model, $\bar{.}$ indicates the mean. The t-test with $n-2$ degrees of freedom is applied.


\subsubsection{Difference of means test}

The null hypothesis is that the means of the two groups are equal. 
The test statistic is $t = \frac{E(y|s^0) - E(y|s^1)}{\sigma \sqrt{1/n_0 + 1/n_1}}$, where $n_0$ is the number of individuals in the regular group, $n_1$ is the number of individuals in the protected group, 
$\sigma = \sqrt{((n_0 - 1)\delta_0^2 + (n_1 - 1)\delta_1^2)/(n_0 + n_1 - 2)}$, where $\delta_0^2$ and  $\delta_1^2$ are the sample target variances in the respective groups. 
The t-test with $n_0 - n_1 -2$ degrees of freedom is applied.

The test assumes independent samples, normality and equal variances. 

\subsubsection{Difference in proportions for two groups}

The null hypothesis is that the rates of positive outcomes within the two groups are equal. 
The test statistic is\\ $z = \frac{p(y^+|s^0) - p(y^+|s^0)}{\sigma}$, where $\sigma  = \sqrt{\frac{p(y^+|s^0)p(y^-|s^0)}{n_0} + \frac{p(y^+|s^1)p(y^-|s^1)}{n_1}}$. The z-test is used.

\subsubsection{Difference in proportions for many groups}

The null hypothesis is that the probabilities or proportions are equal for all the groups. This can be used for testing many groups at once. For example, equality of decisions for different ethnic groups, or age groups. If the null hypothesis is rejected that means at least one of the groups has statistically significantly different proportion. 
The text statistic is\\ $\chi^2 = \sum_{i=1}^k \frac{(n_i- np(y^+|s^i))^2}{p(y^+|s^i)}$, where $k$ is the number of groups. The Chi-Square test is used with $k-1$ degrees of freedom.

\subsubsection{Other tests and related fields}

Relation to clinical trials where protected attribute is the treatment, and outcome is recovery. Prove that there is an effect (there is a discrimination). Does not prove that there is no discrimination. Neither say anything about the magnitude. For example, reduce the flue recovery by 10 min. (practically irrelevant). It may be still relevant for discrimination. Also marketing (measuring the effects of intervention). 

\textbf{Rank test}
MannÐWhitney U test is applied for comparing two groups when the normality and equal variances assumptions are not satisfied. The null hypothesis is that the distributions of the two populations are identical. 
The procedure is to rank all the observations from the largest $y$ to the smallest. The test statistic is the sum of ranks of the protected group. 

\subsection{Absolute measures}
\label{sec:absolutemeasures}

Absolute measures are designed to capture the magnitude of the differences between (typically two) groups of people. The groups are determined by the protected characteristic (e.g. one group is males, another group is females). 
If more than one protected group is analyzed (e.g. different nationalities), typically each group is compared separately to the most favored group.

\subsubsection{Mean difference} Mean difference measures the difference between the means of the targets of the protected group and the general group, $d = E(y^+|s^0) - E(y^+|s^1)$.
If there is not difference then it is considered that there is no discrimination. The measure relates to the difference of means, and difference in proportions test statistics, except that there is no correction for the standard deviation. 

The mean difference for binary classification with binary protected feature, $d = p(y^+|s^0) - p(y^+|s^1)$, is also known as the discrimination score \cite{Calders10}, or \emph{sliftd} \cite{Pedrechi09}. 

Mean difference has been the most popular measure in early work on  non-discriminatory machine learning and data mining \cite{Pedrechi09,Calders10,Kamiran09,Kamiran10,Calders13,Zemel13}. 


\subsubsection{Normalized difference} Normalized difference \cite{Zliobaite15} is the mean difference for binary classification normalized by the rate of positive outcomes, 
 $\delta = \frac{p(y^+|s^0) - p(y^+|s^1)}{d_{\mathit{max}}}$, where $d_{\mathit{max}} =  \min \left(\frac{p(y^+)}{p(s^0)},\frac{p(y^-)}{p(s^1)} \right)$. 
 This measure takes into account maximum possible discrimination at a given positive outcome rate, such that with maximum possible discrimination  at this rate $\delta = 1$, while $\delta = 0$ indicates no discrimination. 

\subsubsection{Area under curve (AUC)} This measure relates to rank tests. It has been used in \cite{Calders13} for measuring discrimination between two groups when the target variable is numeric (regression task), $AUC = \frac{\sum_{(s^i,y^i) \in D^0} \sum_{(s^j,y^j) \in D^1}{\bf I}(y_i > y_j)}{n_0n_1}$, where ${\bf I}(true) = 1$ and $0$ otherwise.

For large datasets computation becomes time and memory intensive, since a quadratic number of comparisons to the number of observations is required. The authors did not mention, but there is an alternative way to compute based on ranking, which, depending on the speed ranking algorithm, may be faster. 
Assign numeric ranks to all the observations, beginning with 1 for the smallest value. Let $R_0$ be the sum of the ranks for the favored group. Then $AUC = R_0 - \frac{n_0(n_0 + 1)}{2}$. 

We observe that if the target variable is binary, and in case of equality half of a point is added to the sum, then AUC linearly relates to mean difference as\\ 
$AUC = p(y^+|s^0)p(y^-|s^1) + 0.5p(y^+|s^0)p(y^+|s^1) + 0.5p(y^-|s^0)p(y^-|s^0) = 0.5d + 0.5$, where $d$ denotes discrimination measured by the mean difference measure. 

\subsubsection{Impact ratio} Impact ratio, also known as slift \cite{Pedrechi09}, is the ratio of positive outcomes for the protected group over the general group, $r = p(y^+|s^1)/p(y^+|s^0)$.
This measure is used in the US courts for quantifying discrimination, 
the decisions are deemed to be discriminatory if the ratio of positive outcomes for the protected group is below $80\%$ of that of the general group. 
Also this is the form stated in the Sex Discrimination Act of U.K. $r=1$ indicates that there is no discrimination.

\subsubsection{Elift ratio} Elift ratio \cite{Pedrechi08} is similar to impact ratio, but instead of  dividing by the general group, 
the denominator is the overall rate of positive outcomes $r = p(y^+|s^0)/p(y^+)$. The same measure, expressed as $p\frac{p(y,s)}{p(y)p(s)} < 1 + \eta$ for all values of $y$ and $s$, is later referred to as $\eta$-neutrality \cite{Kamishima13}. 

\subsubsection{Odds ratio} Odds ratio of two proportions is often used in natural, social and biomedical sciences to measure the association between exposure and outcome. The popularity is due to convenient  relation with the logistic regression. The exponential function of the logistic regression coefficient translates one unit increase in the odds ratio. 
Odds ratio has been used for measuring discrimination \cite{Pedrechi09} as $r = \frac{p(y^+|s^0)p(y^-|s^1)}{p(y^+|s^1)p(y^-|s^0)}$.

\subsubsection{Mutual information} Mutual information (MI) is popular in information theory for measuring mutual dependence between variables. In discrimination literature this measure has been referred to as normalized prejudice index \cite{Kamishima13}, and used for measuring the magnitude of discrimination. 
Mutual information is measured in bits, but it can be normalized such that the result falls into the range between $0$ and $1$. For categorical variables $MI = \frac{I(y,s)}{\sqrt{H(y),H(s)}}$, where $I(s,y) = \sum_{(s,y)} p(s,y) \log \frac{p(s,y)}{p(s)p(y)}$, and $H(y) = - \sum_{y} p(y) \log p(y)$.
For numerical variables the summation is replaces by integral.

\subsubsection{Balanced residuals} While other measures work on datasets, balanced residuals is for machine learning model outputs. This measure characterizes the difference between the actual outcomes recorded in the dataset, and the model outputs. The requirement is that underpredictions and overpredictions should be balanced within the protected and regular groups. \cite{Calders13} proposed balanced residuals as a criteria, not a measure. That is, the average residuals should be equal, but in principle the difference could be used as a measure of discrimination  $d = \frac{\sum_{i \in D^1} y_i - \hat{y}_i}{n_1} - \frac{\sum_{j \in D^0} y_j - \hat{y}_j}{n_0}$, where $y$ is the true target value, $\hat{y}$ is the prediction. Positive values of $d$ would indicate discrimination towards the protected group. 
One should; however, use and interpret this measure with caution. If the learning dataset is discriminatory, but the predictive model makes ideal predictions such that all the residuals are zero, this measure would show no discrimination, even though the predictions would be discriminatory, since the original data is discriminatory. Suppose, another predictive model makes a constant prediction for everybody, and the constant prediction is equal to the mean of the regular group. If the learning dataset contains discrimination, then the residuals for the regular group would be smaller than for the protected group, and the measure would indicate discrimination, however, a constant prediction to everybody means tat everybody is treated equally, and there should be no discrimination detected.

\subsubsection{Other possible measures}

There are many established measures in feature selection literature \cite{Guyon03} for measuring the relation between two variables, which, in principle, can be used as absolute discrimination measures. The stronger the relation between the protected variable $s$ and the target variable $y$, the larger the absolute discrimination. 

There are three main groups of measures for relation between variables: correlation based, information theoretic, and one-class classifiers. Correlation based measures, such as the Person correlation coefficient, are typically used for numeric variables. Information theoretic measures, such as mutual information mentioned earlier, are typically used for categorical variables. One-class classifiers present an interesting option. In discrimination the setting would be to predict the target $y$ solely on the protected variable $s$, and measure the prediction accuracy. 
We are not aware of such attempts in the non-discriminatory machine learning literature, but it would be a valid option to explore. 

\subsubsection{Measuring for more than two groups}

Most of the absolute discrimination measures are for two groups (protected group vs. regular group). Ideas, how to apply those for more than two groups, can be borrowed from multi-class classification \cite{Bishop06_book}, multi-label classification \cite{Tsoumakas07}, and one-class classification \cite{Tax01} literature. 
Basically, there are three options how to obtain sub-measures: measure pairwise for each pair of groups ($k(k-1)/2$ comparisons), measure one against the rest for each group ($k$ comparisons), measure each group against the regular group ($k-1$ comparisons). The remaining question is how to aggregate the sub-measures. Based on personal conversations with legal experts, we advocate for reporting the maximum from all the comparisons as the final discrimination score. Alternatively, all the scores could be summed weighing by the group sizes to obtain an overall discrimination score. 





Even though absolute measures do not take into account any explanations of possible differences of decisions across groups, they can be considered as core building blocks for developing conditional measures. Conditional measures do take into account explanations in differences, and measure only discrimination that cannot be explained by non-protected characteristics.

Table \ref{tab:tasks} summarizes applicability of absolute measures in different machine learning settings. 
\begin{table}
\tbl{Summary of absolute measures. Checkmark (\checkmark) indicates that it is directly applicable in a given machine learning setting. 
Tilde ($\sim$) indicates that a straightforward extension exists (for instance, measuring pairwise). 
\label{tab:tasks}}{%
{\small
\begin{tabular}{lcccccc}
\toprule
& \multicolumn{3}{c}{Protected variable} & \multicolumn{3}{c}{Target variable}\\
\cmidrule(r){2-4}
\cmidrule(r){5-7} 
Measure & Binary & Categoric & Numeric & Binary & Ordinal & Numeric \\
\midrule
Mean difference & \checkmark & $\sim$ & & \checkmark &  & \checkmark \\
Normalized difference & \checkmark & $\sim$ & & \checkmark &  & \\
Area under curve & \checkmark & $\sim$ & & \checkmark &  \checkmark &  \checkmark \\
Impact ratio & \checkmark & $\sim$ & & \checkmark &  & \\
Elift ratio & \checkmark & $\sim$ & & \checkmark &  & \\
Odds ratio & \checkmark & $\sim$ & & \checkmark &  & \\
Mutual information & \checkmark & \checkmark & \checkmark & \checkmark & \checkmark & \checkmark \\
Balanced residuals & \checkmark & $\sim$ & & $\sim$ & \checkmark & \checkmark \\
Correlation & \checkmark & & \checkmark & \checkmark & & \checkmark \\
\bottomrule
\end{tabular}
}}
\end{table}

\subsection{Conditional measures}
\label{sec:conditionalmeasures}

Absolute measures take into account only the target variable $y$ and the protected variable $s$.
Absolute measures consider all the differences in treatment between the protected group and the regular group to be discriminatory. 
Conditional measure, on the other hand, try to capture how much of the difference between the groups is explainable by other characteristics of individuals, recorded in $X$, and only the remaining differences are deemed to be discriminatory.  For example, part of the difference in acceptance rates for natives and immigrants may be explained by the difference in education level. Only the remaining unexplained difference should be considered as discrimination. Let $z = f(X)$ be an explanatory variable. For example,  if $z^i$ denotes a certain education level. Then all the individuals with the same level of education will form a strata $i$. Within each strata the acceptance rates are required to be equal. 

\subsubsection{Unexplained difference} 
Unexplained difference \cite{Kamiran13} is measured, as the name suggests, as the overall mean difference minus the differences that can be explained by other legitimate variable. Recall that mean difference is $d = p(y^+|s^0) - p(y^+|s^1)$. 
Then the unexplained difference $d_u = d - d_e$, 
where\\ $d_e = \sum_{i=1}^m p^\star(y^+|z^i)(p(z^i|s^0) -  p(z^i|s^1))$, where  $p^\star(y^+|z^i)$ is the desired acceptance rate within the strata $i$. 
The authors recommend using
$p^\star(y^+|z^i) = \frac{p(y^+|s^0,z^i) + p(y^+|s^1,z^i)}{2}$. In the simplest case $z$ bay be equal one of the variables in $X$. The authors also use clustering on $X$ to take into account more than one explanatory variable at the same time. Then $z$ denotes a cluster, one strata is one cluster. 

\subsubsection{Propensity measure} Propensity models \cite{Rosenbaum83} are typically used in clinical trials or marketing for estimating the probability that an individual would receive a treatment. Given the estimated probabilities, individuals can be stratified according to similar probabilities of receiving a treatment, and the effects of treatment can be measured within each strata separately. 
Propensity models have been used for measuring discrimination \cite{Calders13}, in this case a function was learned to model the protected characteristic based on input variables $X$, that is $s^1 = f(X)$. A logistic regression was used for modeling $f(.)$. Then the estimated propensity scores $\hat{s}^1$ were split into five ranges, where each range formed one strata. Discrimination was measured within each strata, treating each strata as a separate dataset, and using absolute discrimination measures discussed in the previous section. The authors did not aggregate the resulting discrimination into one measure, but in principle the results can be aggregated into one measure, for instance, using the unexplained difference formulas, reported above. In such a case each strata would correspond to one value of an explanatory variable $z$.

\subsubsection{Belift ratio} Belift ratio \cite{Mancuhan14} is similar to Elift ratio in absolute measures, but here the probabilities of positive outcome are also conditioned on input attributes, $belift = \frac{p(y^+|s^1,X^r,X^a)}{p(y^+|X^a)}$, where $X = X^r \cup X^{\not r}$ is a set of input variables, $X^{r}$ denotes so caller redlining attributes, the variables which are correlated with the protected variable $s$. The authors proposed estimating the probabilities via bayesian networks. A possible difficulty for applying this measure in practice may be that not everybody, especially non-machine learning users, are familiar enough with the Bayesian networks to an extent needed for estimating the probabilities. Moreover, construction of a Bayesian network may be different even for the same problem depending on assumptions made about interactions between the variables. Thus, different users may get different discrimination scores for  the same application case.

A simplified approximation of belift could be to treat all the attributes as redlining attributes, and instead of conditioning on all the input variables, condition on a summary of input variables $z$, where $z = f(X)$. Then the measure for strata $i$ would be $\frac{p(y^+|s^1,z^i)}{p(y^+)}$. 

The measure has a limitation that neither the original version, nor the simplified version allow differences to be explained by variables that are correlated with the protected variable. That is, if a university has two programmes, say medicine and computer science, and the protected group, e.g. females, are more likely to apply for a more competitive programme, then the programmes cannot have different acceptance rates. That is, if the acceptance rates are different, all the difference is considered to discriminatory. 

\subsection{Structural measures}

Structural measures are targeted at quantifying direct discrimination. The main idea behind structural measures is for each individual in the dataset to identify whether s/he is discriminated, and then analyze how many individuals in the dataset are affected. Currently 

\subsubsection{Situation testing}
Situation testing \cite{Luong11} measures which fraction of individuals in the protected group are considered discriminated, as 
$f =  \frac{\sum_{y_i \in D(y^0|s^1)} {\bf I}(\mathit{diff}(y_i) \geq t)}{|D(y^0|s^1)|}$, where $t$ is a user defined threshold, ${\bf I}$ is the indicator function that takes $1$ if true, $0$ otherwise. 
The situation testing for an individual $i$ is computed  as $\mathit{diff}(y_i) = \frac{\sum_{y_j \in D^0{\mathit{\kappa-nearest-neighbours} }}}{\kappa} - \frac{\sum_{y_j \in D^1{\mathit{\kappa-nearest-neighbours} }}}{\kappa}$.
Positive and negative discrimination is handled separately.

The idea is to compare each individual to the opposite group and see if the decision would be different. 
In that sense, the measure relates to propensity scoring (Section \ref{sec:conditionalmeasures}), used for identifying groups of people similar according to the  non-protected characteristics, and requiring for decisions within those groups to be balanced. The main difference is that propensity measures would signal indirect discrimination within a group, and situation testing aims at signalling direct discrimination for each individual in question.

\subsubsection{Consistency} Consistency measure \cite{Zemel13} compares the predictions for each individual with his/her nearest neighbors. 
$C = 1 - \frac{1}{\kappa N}\sum_{i=1}^N \sum_{y_j \in D^{\mathit{\kappa-nearest-neighbours}}}|y_i - y_j|$.
Consistency measure is closely related to situation testing, but considers nearest neighbors from any group (not from the opposite group).  
Due to this choice, consistency measure should be used with caution in situations where there is a high correlation between the protected variable and the legitimate input variables. For example, suppose we have only one predictor variable - location of an apartment, and the target variable is to grant a loan or not. Suppose all non-white people live in one neighborhood (as in the redlining example), and all the white people in the other neighborhood. Unless the number of nearest neighbors to consider is very large, this measure will show no discrimination, since all the neighbors will get the same decision, even though all black residents will be rejected, and all white will be accepted (maximum discrimination). Perfect consistency, but maximum discrimination. In their experimental evaluation the authors have used this measure in combination with the mean difference measure. 

%
%
%
%
%


\section{Analysis of core measures}
\label{sec:analysis}

Even though absolute measures are naive  in a sense that they do not take any possible explanations of different treatment into account, and due to that may show more discrimination that there actually is, these measures provide core mechanisms and a basis for measuring indirect discrimination. Conditional measures are typically built upon absolute measures. In addition, statistical tests often directly relate to absolute measures. 
Thus, to provide a better understanding of properties and implications of choosing one measure over another, in this section we computationally analyze a set of absolute measures, and discuss their properties. 

We analyze the following measures, introduced in Section \ref{sec:absolutemeasures}: mean difference, normalized difference, mutual information, impact ratio, elift and odds ratio. From the measures analyzed in this section, mean difference and area under curve can be directly used in regression tasks. We focus on the classification scenario, since this scenario has been studied more extensively in the discrimination-aware data mining and machine learning literature, and there are more measures available for classification than for regression; the regression setting, except for a recent work \cite{Calders13}, remains a subject of future research, and therefore is out of the scope of a survey paper. 

Table \ref{tab:limits} summarizes boundary conditions of the selected measures. In the difference based measures $0$ indicates no discrimination, in the ratio based measures $1$ indicates no discrimination, in AUC $0.5$ means no discrimination. The boundary conditions are reached when one group gets all the positive  decisions, and the other group gets all the negative decisions. 
\begin{table}
\tbl{Limits\label{tab:limits}}{%
\begin{tabular}{lccc}
\hline
Measure & Maximum & No & Reverse \\
& discrimination & discrimination & discrimination \\
\hline
\hline
\multicolumn{4}{l}{Differences}\\
\hline 
Mean difference & $1$ & $0$ & $-1$ \\
Normalized difference & $1$ & $0$ & $-1$ \\
Mutual information & $1$ & $0$ & $1$ \\
\hline
\hline
\multicolumn{4}{l}{Ratios}\\
\hline
Impact ratio & $0$ & $1$ & $+\infty$ \\
Elift & $0$ & $1$ & $+\infty$ \\
Odds ratio & $0$ & $1$ & $+\infty$ \\
\hline
\hline
\multicolumn{4}{l}{AUC}\\
\hline
Area under curve (AUC) & $1$ & $0.5$ & $0$ \\
\hline
\hline
\end{tabular}
}
\end{table}



Next we experimentally analyze the performance of the selected measures. We leave out AUC from the experiments, since in classification it is equivalent to the mean difference measure. The goal of the experiments is to demonstrate how the performance depends on variations in the overall rate of positive decisions, balance between classes and balance between the regular and protected groups of people in data.  

For this analysis we use synthetically generated data which allows to represent different task settings and control the levels of underlying discrimination.
Given four parameters: 
the proportion of individuals in the protected group $p(s^1)$, 
the proportion of positive outputs $p(y^+)$, the underlying discrimination $d \in [-100\%,100\%]$, and the number of data points $n$, data is generated as follows.
First $n$ data points are generated assigning a score in $[0,1]$ uniformly at random, and assigning group membership at random according to the probability $p(s^1)$. This data contains no discrimination, because the scores are assigned at random. If would contain full discrimination if we ranked the observations according to the assigned scores and all the members of the regular group would appear before all the members of the protected group. Following this reasoning, half-discrimination would be if in a half of the data the members of the regular group appear before all the members of the protected group in the ranking, and the other half of the data would show a random mix of both groups in the ranking. For the experimental analysis purposes we define this as $50\%$ discrimination. It is difficult to measure discrimination in data this way, but it is easy to generate such a data. For a given level of desired discrimination $d$ we select $dn$ observations at random, sort them according to their scores, and then permute group assignments within this subsample in such a way that the highest scores get assigned to the regular group, and the lowest scores get assigned to the protected group. Finally, since the experiment is about classification, we round the scores to zero-one in such a way that the proportion of ones is as desired by $p(y^+)$. Then we apply different measures of discrimination to data generated this way, and investigate, how these measures can reconstruct the underlying discrimination. For each parameter setting we generate $n=10000$ data points, and average the results over $100$ such runs\footnote{The code for our experiments is made available at \url{https://github.com/zliobaite/paper-fairml-survey}.}

Figure \ref{fig:differences} depicts  the performance of mean difference, normalized difference and mutual information. Ideally, the performance should be invariant to balance of the groups ($p(s^10)$) and the proportion of positive outputs ($p(y^+)$), and thus run along the diagonal line in as many plots, as possible. We can see that the normalized difference captures that. The mean difference captures the trends, but the indicated discrimination highly depends on the balance of the classes and balance of the groups, therefore, this measure to be  interpreted with care when data is highly imbalanced. The same holds for mutual information. For instance, at $p(s^1)= 90\%$ and $p(y^+)=90\%$ the true discrimination in data may be near $100\%$, i.e. nearly the worst possible, but both measures would indicate that discrimination is nearly zero. The normalized difference would capture the situation as desired. In addition to that, we see that the mean difference and normalized difference are linear measures, while mutual information is non-linear, and would show less discrimination that actually in the medium ranges. Moreover, mutual information dos not indicate the sign of discrimination, that is, the outcome does not indicate whether discrimination is reversed or not. For these reasons, we do not recommend using mutual information for the purpose of quantifying discrimination.
Therefore, from the difference based measures we advocate normalized difference, which was designed to be robust to imbalances in data. 
The normalized difference  is somewhat more complex to compute than the mean difference, which may be a limitation for practical applications outside research.
Therefore, if data is closed to balanced in terms of groups and  positive-negative outputs, then the mean difference can be used. 

\newcommand{\plotw}{4.5cm}
\newcommand{\ploth}{4.5cm}
\pgfplotsset{ylabel right/.style={after end axis/.append code={\node [rotate=90, anchor=north] at (rel axis cs:1,0.5) {#1};}}}
\begin{figure}
\newcommand{\plotaxis}{width = \plotw,height=\ploth, ymin = -1.1, ymax = 1.1,xmin = -1.1, xmax = 1.1, xtick={-1,0,1}, ytick = {-1,0,1}}
\newcommand{\plotplot}{
\addplot[black,domain=-1:1, samples=3,dotted,line width = 0.5pt]{x*0};
\addplot[black,domain=-1:1, samples=3,dotted,line width = 0.5pt] coordinates {(0,-1)(0,1)};
\addplot[mycol2,line width = 2.5pt]  table[x = disc, y = ddif] {\filenow};
\addplot[mycol5,line width = 3.5pt, dashed]  table[x = disc, y = ddnorm] {\filenow};
\addplot[mycol3,line width = 1.5pt]  table[x = disc, y = MInorm] {\filenow};
}
\ref{named}\\
\begin{tikzpicture}[scale=0.7]
\newcommand{\filenow}{out_pF10_pp10.dat}
\begin{axis}[name = plot1,\plotaxis,ylabel = measured discrim.,legend entries={,,mean difference, normalized difference, mutual information},
      legend style={draw=none,font=\scriptsize},legend columns=3,legend to name = named,title = {$p(y^+) = 10\%$}] \plotplot \end{axis}
\renewcommand{\filenow}{out_pF10_pp30.dat}
\begin{axis}[name = plot2, at=(plot1.right of south east), anchor=left of south west,\plotaxis,title = {$30\%$}] \plotplot \end{axis}
\renewcommand{\filenow}{out_pF10_pp50.dat}
\begin{axis}[name = plot3, at=(plot2.right of south east), anchor=left of south west, \plotaxis,title = {$50\%$}] \plotplot \end{axis}
\renewcommand{\filenow}{out_pF10_pp70.dat}
\begin{axis}[name = plot4, at=(plot3.right of south east), anchor=left of south west, \plotaxis,title = {$70\%$}] \plotplot \end{axis}
\renewcommand{\filenow}{out_pF10_pp90.dat}
\begin{axis}[name = plot5, at=(plot4.right of south east), anchor=left of south west, \plotaxis,title = {$90\%$},ylabel right = {$p(s^1) = 10\%$}] \plotplot \end{axis}
\renewcommand{\filenow}{out_pF30_pp10.dat}
\begin{axis}[name=plot6, at=(plot1.below south west), anchor=above north west, \plotaxis,ylabel = measured discrim.] \plotplot \end{axis}
\renewcommand{\filenow}{out_pF30_pp30.dat}
\begin{axis}[name=plot7, at=(plot6.right of south east), anchor=left of south west,\plotaxis] \plotplot \end{axis}
\renewcommand{\filenow}{out_pF30_pp50.dat}
\begin{axis}[name=plot8, at=(plot7.right of south east), anchor=left of south west,\plotaxis] \plotplot \end{axis}
\renewcommand{\filenow}{out_pF30_pp70.dat}
\begin{axis}[name=plot9, at=(plot8.right of south east), anchor=left of south west,\plotaxis] \plotplot \end{axis}
\renewcommand{\filenow}{out_pF30_pp90.dat}
\begin{axis}[name=plot10, at=(plot9.right of south east), anchor=left of south west,\plotaxis,ylabel right = {$30\%$}] \plotplot \end{axis}
\renewcommand{\filenow}{out_pF50_pp10.dat}
\begin{axis}[name=plot11, at=(plot6.below south west), anchor=above north west, \plotaxis,ylabel = measured discrim.] \plotplot \end{axis}
\renewcommand{\filenow}{out_pF50_pp30.dat}
\begin{axis}[name=plot12, at=(plot11.right of south east), anchor=left of south west,\plotaxis] \plotplot \end{axis}
\renewcommand{\filenow}{out_pF50_pp50.dat}
\begin{axis}[name=plot13, at=(plot12.right of south east), anchor=left of south west,\plotaxis] \plotplot \end{axis}
\renewcommand{\filenow}{out_pF50_pp70.dat}
\begin{axis}[name=plot14, at=(plot13.right of south east), anchor=left of south west,\plotaxis] \plotplot \end{axis}
\renewcommand{\filenow}{out_pF50_pp90.dat}
\begin{axis}[name=plot15, at=(plot14.right of south east), anchor=left of south west,\plotaxis,ylabel right = {$50\%$}] \plotplot \end{axis}
\renewcommand{\filenow}{out_pF70_pp10.dat}
\begin{axis}[name=plot16, at=(plot11.below south west), anchor=above north west, \plotaxis,ylabel = measured discrim.] \plotplot \end{axis}
\renewcommand{\filenow}{out_pF70_pp30.dat}
\begin{axis}[name=plot17, at=(plot16.right of south east), anchor=left of south west,\plotaxis] \plotplot \end{axis}
\renewcommand{\filenow}{out_pF70_pp50.dat}
\begin{axis}[name=plot18, at=(plot17.right of south east), anchor=left of south west,\plotaxis] \plotplot \end{axis}
\renewcommand{\filenow}{out_pF70_pp70.dat}
\begin{axis}[name=plot19, at=(plot18.right of south east), anchor=left of south west,\plotaxis] \plotplot \end{axis}
\renewcommand{\filenow}{out_pF70_pp90.dat}
\begin{axis}[name=plot20, at=(plot19.right of south east), anchor=left of south west,\plotaxis,ylabel right = {$70\%$}] \plotplot \end{axis}
\renewcommand{\filenow}{out_pF90_pp10.dat}
\begin{axis}[name=plot21, at=(plot16.below south west), anchor=above north west, \plotaxis,ylabel = measured discrim.,xlabel = discrim. in data] \plotplot \end{axis}
\renewcommand{\filenow}{out_pF90_pp30.dat}
\begin{axis}[name=plot22, at=(plot21.right of south east), anchor=left of south west,\plotaxis, xlabel = discrim. in data] \plotplot \end{axis}
\renewcommand{\filenow}{out_pF90_pp50.dat}
\begin{axis}[name=plot23, at=(plot22.right of south east), anchor=left of south west,\plotaxis, xlabel = discrim. in data] \plotplot \end{axis}
\renewcommand{\filenow}{out_pF90_pp70.dat}
\begin{axis}[name=plot24, at=(plot23.right of south east), anchor=left of south west,\plotaxis, xlabel = discrim. in data] \plotplot \end{axis}
\renewcommand{\filenow}{out_pF90_pp90.dat}
\begin{axis}[name=plot24, at=(plot24.right of south east), anchor=left of south west,\plotaxis, xlabel = discrim. in data,ylabel right = {$90\%$}] \plotplot \end{axis}
\end{tikzpicture}
\caption{Analysis of the measures based on differences: discrimination in data vs. measured discrimination.}
\label{fig:differences}
\end{figure}
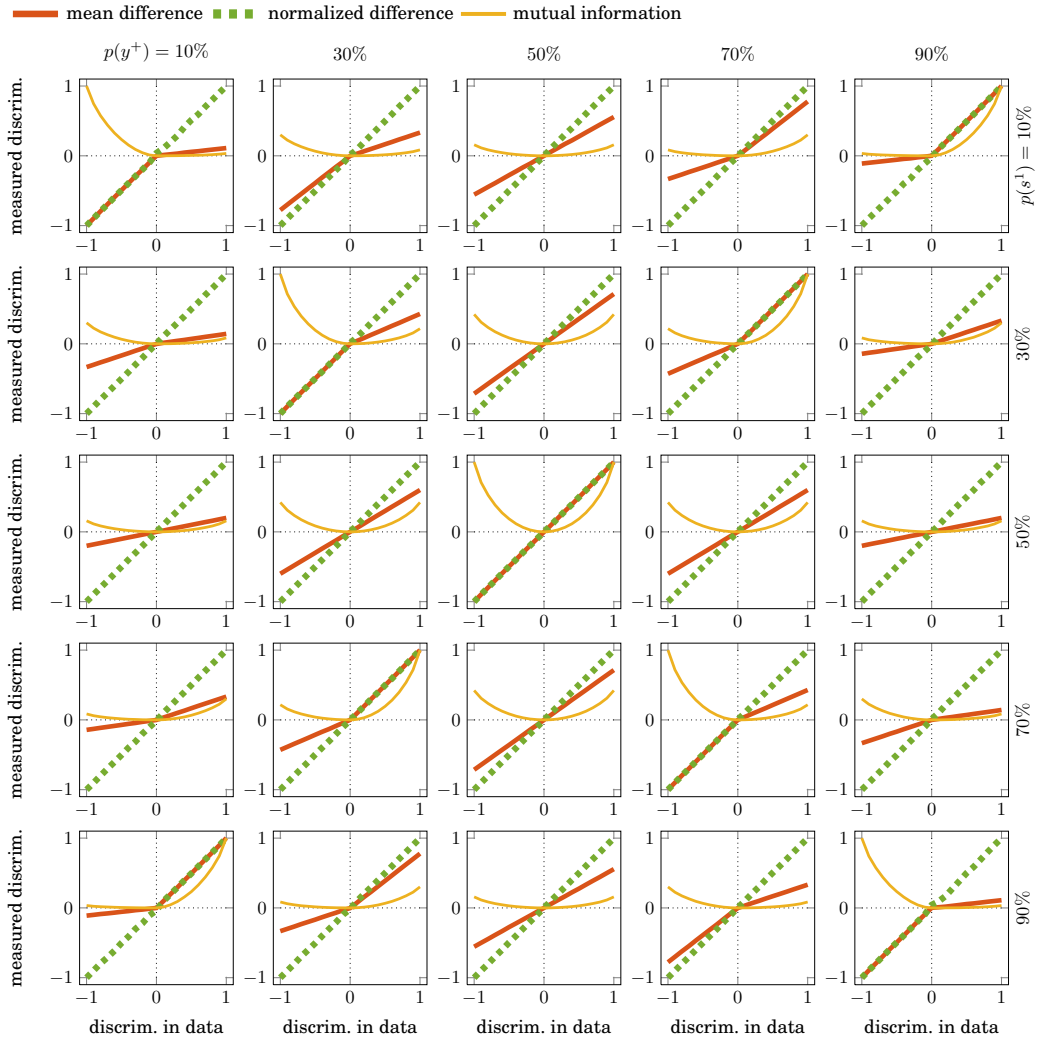

%

\begin{figure}
\newcommand{\plotaxis}{width = \plotw,height=\ploth,  ymin = -0.1, ymax = 3.1,xmin = -1.1, xmax = 1.1, xtick={-1,0,1}, ytick = {0,1,2,3}}
\newcommand{\plotplot}{
\addplot[black,domain=-1:1, samples=3,dotted,line width = 0.5pt]{x*0+1};
\addplot[black,domain=-1:1, samples=3,dotted,line width = 0.5pt] coordinates {(0,-1)(0,3)};
\addplot[mycol6,line width = 2.5pt]  table[x = disc, y = dratio] {\filenow};
\addplot[mycol1,line width = 3.5pt, dashed]  table[x = disc, y = dolift] {\filenow};
\addplot[mycol4,line width = 1.5pt]  table[x = disc, y = delift] {\filenow};
}
\ref{named2}\\
\begin{tikzpicture}[scale=0.7]
\newcommand{\filenow}{out_pF10_pp10.dat}
\begin{axis}[name = plot1,\plotaxis,ylabel = measured discrim.,legend entries={,,impact ratio, elift, odds ratio},
      legend style={draw=none,font=\scriptsize},legend columns=3,legend to name = named2,title = {$p(y^+) = 10\%$}] \plotplot \end{axis}
\renewcommand{\filenow}{out_pF10_pp30.dat}
\begin{axis}[name = plot2, at=(plot1.right of south east), anchor=left of south west,\plotaxis,title = {$30\%$}] \plotplot \end{axis}
\renewcommand{\filenow}{out_pF10_pp50.dat}
\begin{axis}[name = plot3, at=(plot2.right of south east), anchor=left of south west, \plotaxis,title = {$50\%$}] \plotplot \end{axis}
\renewcommand{\filenow}{out_pF10_pp70.dat}
\begin{axis}[name = plot4, at=(plot3.right of south east), anchor=left of south west, \plotaxis,title = {$70\%$}] \plotplot \end{axis}
\renewcommand{\filenow}{out_pF10_pp90.dat}
\begin{axis}[name = plot5, at=(plot4.right of south east), anchor=left of south west, \plotaxis,title = {$90\%$},ylabel right = {$p(s^1) = 10\%$}] \plotplot \end{axis}
\renewcommand{\filenow}{out_pF30_pp10.dat}
\begin{axis}[name=plot6, at=(plot1.below south west), anchor=above north west, \plotaxis,ylabel = measured discrim.] \plotplot \end{axis}
\renewcommand{\filenow}{out_pF30_pp30.dat}
\begin{axis}[name=plot7, at=(plot6.right of south east), anchor=left of south west,\plotaxis] \plotplot \end{axis}
\renewcommand{\filenow}{out_pF30_pp50.dat}
\begin{axis}[name=plot8, at=(plot7.right of south east), anchor=left of south west,\plotaxis] \plotplot \end{axis}
\renewcommand{\filenow}{out_pF30_pp70.dat}
\begin{axis}[name=plot9, at=(plot8.right of south east), anchor=left of south west,\plotaxis] \plotplot \end{axis}
\renewcommand{\filenow}{out_pF30_pp90.dat}
\begin{axis}[name=plot10, at=(plot9.right of south east), anchor=left of south west,\plotaxis,ylabel right = {$30\%$}] \plotplot \end{axis}
\renewcommand{\filenow}{out_pF50_pp10.dat}
\begin{axis}[name=plot11, at=(plot6.below south west), anchor=above north west, \plotaxis,ylabel = measured discrim.] \plotplot \end{axis}
\renewcommand{\filenow}{out_pF50_pp30.dat}
\begin{axis}[name=plot12, at=(plot11.right of south east), anchor=left of south west,\plotaxis] \plotplot \end{axis}
\renewcommand{\filenow}{out_pF50_pp50.dat}
\begin{axis}[name=plot13, at=(plot12.right of south east), anchor=left of south west,\plotaxis] \plotplot \end{axis}
\renewcommand{\filenow}{out_pF50_pp70.dat}
\begin{axis}[name=plot14, at=(plot13.right of south east), anchor=left of south west,\plotaxis] \plotplot \end{axis}
\renewcommand{\filenow}{out_pF50_pp90.dat}
\begin{axis}[name=plot15, at=(plot14.right of south east), anchor=left of south west,\plotaxis,ylabel right = {$50\%$}] \plotplot \end{axis}
\renewcommand{\filenow}{out_pF70_pp10.dat}
\begin{axis}[name=plot16, at=(plot11.below south west), anchor=above north west, \plotaxis,ylabel = measured discrim.] \plotplot \end{axis}
\renewcommand{\filenow}{out_pF70_pp30.dat}
\begin{axis}[name=plot17, at=(plot16.right of south east), anchor=left of south west,\plotaxis] \plotplot \end{axis}
\renewcommand{\filenow}{out_pF70_pp50.dat}
\begin{axis}[name=plot18, at=(plot17.right of south east), anchor=left of south west,\plotaxis] \plotplot \end{axis}
\renewcommand{\filenow}{out_pF70_pp70.dat}
\begin{axis}[name=plot19, at=(plot18.right of south east), anchor=left of south west,\plotaxis] \plotplot \end{axis}
\renewcommand{\filenow}{out_pF70_pp90.dat}
\begin{axis}[name=plot20, at=(plot19.right of south east), anchor=left of south west,\plotaxis,ylabel right = {$70\%$}] \plotplot \end{axis}
\renewcommand{\filenow}{out_pF90_pp10.dat}
\begin{axis}[name=plot21, at=(plot16.below south west), anchor=above north west, \plotaxis,ylabel = measured discrim.,xlabel = discrim. in data] \plotplot \end{axis}
\renewcommand{\filenow}{out_pF90_pp30.dat}
\begin{axis}[name=plot22, at=(plot21.right of south east), anchor=left of south west,\plotaxis, xlabel = discrim. in data] \plotplot \end{axis}
\renewcommand{\filenow}{out_pF90_pp50.dat}
\begin{axis}[name=plot23, at=(plot22.right of south east), anchor=left of south west,\plotaxis, xlabel = discrim. in data] \plotplot \end{axis}
\renewcommand{\filenow}{out_pF90_pp70.dat}
\begin{axis}[name=plot24, at=(plot23.right of south east), anchor=left of south west,\plotaxis, xlabel = discrim. in data] \plotplot \end{axis}
\renewcommand{\filenow}{out_pF90_pp90.dat}
\begin{axis}[name=plot24, at=(plot24.right of south east), anchor=left of south west,\plotaxis, xlabel = discrim. in data,ylabel right = {$90\%$}] \plotplot \end{axis}
\end{tikzpicture}
\caption{Analysis of the measures based on ratios: discrimination in data vs. measured discrimination.}
\label{fig:ratios}
\end{figure}
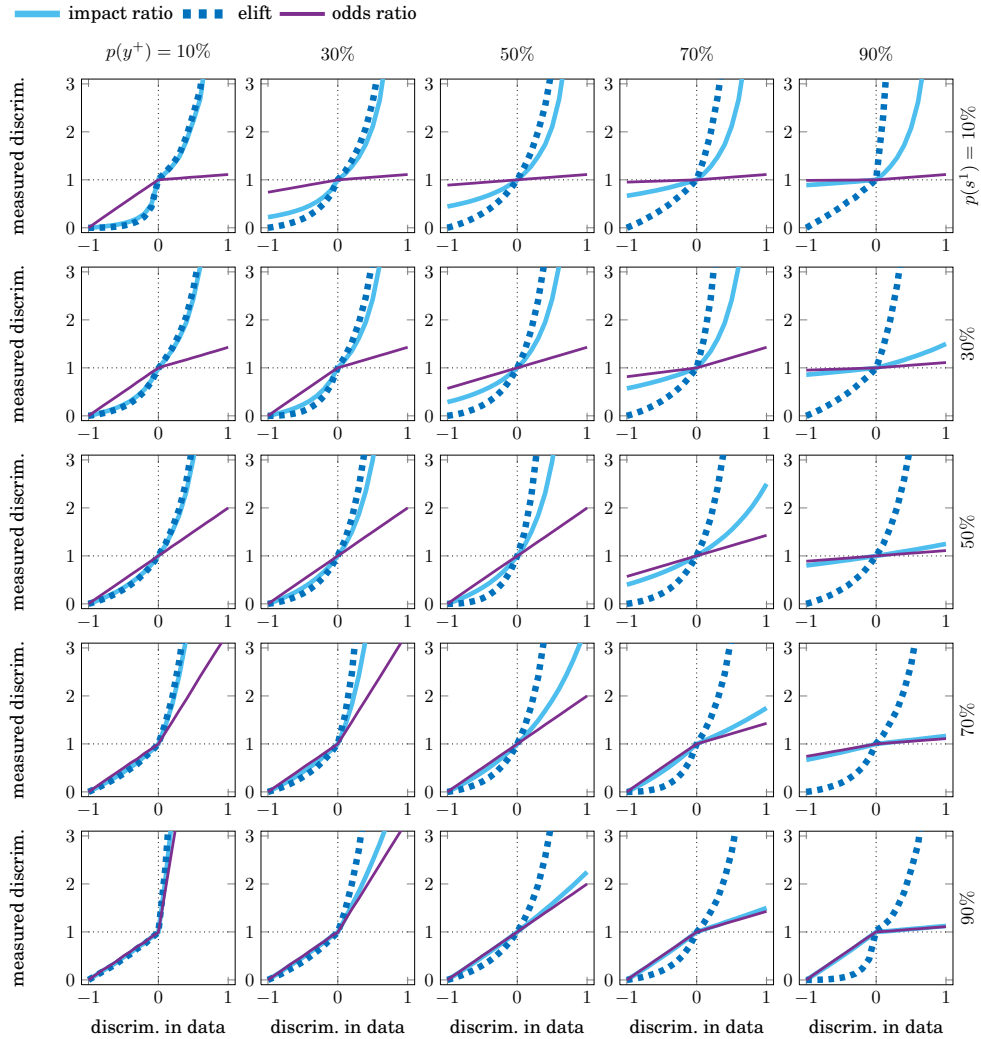

Figure \ref{fig:ratios} presents similar analysis of the measures based on ratios: impact ratio, elift and odds ratio. We can see that the odds ratio, and the impact ratio are very sensitive to imbalances in groups and positive outputs. The elift is more stable in that respect, but still has some variations, particularly at high imbalance of positive outputs ($p(y^+)=90\%$ or $10\%$), when discrimination may be highly exaggerated (far from the diagonal line). 
 In addition, measured discrimination by all ratios grows very fast at low rates of positive outcome (e.g. see the plot $p(y^+) = 10\%$ and $p(s^1)=90\%$), while there is almost no discrimination in the data, measures indicate high discrimination. 
 We also can see that all the ratios are asymmetric in terms of reverse discrimination. One unit of measured discrimination is not the same as one unit of reverse discrimination. This makes ratios a bit more difficult to interpret than differences, analyzed earlier, especially at large scale explorations and comparisons of, for instance, different computational methods for prevention. Due to these reasons, we do not recommend using ratio based discrimination measures, since they are much more difficult to interpret correctly, and may easily be misleading. Instead recommend using and building upon difference based measures, discussed in Figure \ref{fig:differences}.

The core measures that we have analyzed form a basis for assessing fairness of predictive models, but it is not enough to use them directly, since they do not take into account possible legitimate explanations of differences between the groups, and instead consider any differences between the groups of people undesirable. The basic principle is to try to stratify the population in such a way that in each stratum contains people that are similar in terms of their legitimate characteristics, for instance, have similar qualifications if the task is candidate selection for job interviews. 
propensity score matching, reported in Section \ref{sec:conditionalmeasures}, is one possible way to stratification, but it is not the only one, and outcomes may vary depending on internal parameter choices. Thus, the principle to measuring is available, but there are still open challenges ahead to make the approach more robust to different users, and more uniform across different task setting, such that one could diagnose potential discrimination or declare fairness with more confidence. 

\section{Recommendations for researchers and practitioners}
\label{sec:conclusion}

As attention of researchers, media and general public to potential discrimination is growing, it is important to be able to measure fairness of predictive models in a systematic and accountable way. We have surveyed measures used (and potentially usable) for measuring indirect discrimination in machine learning, and experimentally analyzed the performance of the core measures in classification tasks. Based on our analysis we generally recommend using the normalized difference, and in case the classes and groups of people in the data are well balanced, it may be sufficient to use the simple  (unnormalized) mean difference. We do not recommend using ratio based measures challenges associated with their interpretation in different situation. 

The core measures stand alone are not enough for measuring fairness correctly. These measures can only be applied to uniform populations considering that everybody within the population is equally qualified to get a positive decision. In reality this is rarely the case, for example, different salary levels may be explained by different education levels. Therefore, the main principle of applying the core measures should be by first segmenting the population into more or less uniform segments according to their qualifications, and then applying core measures within each segment. Some of such measuring techniques have been surveyed in Section \ref{sec:conditionalmeasures} (Conditional measures), but generally there is no one easy way to approach it, and presenting sound arguments to justify the methods of allocating people into segments is very important in research and practice. 

We hope that this survey can establish a basis for further research developments in this important topic. So far most of the research has concentrated on binary classification with binary protected characteristic. While this is a base scenario, relatively easy to deal with in research, many technical challenges for future research lie in addressing more complex learning scenarios with different types and multiple protected characteristics, in multi-class, multi-target classification and regression settings, with different types of legitimate variables, noisy input data, potentially missing protected characteristics, and many more. 

\bibliographystyle{ACM-Reference-Format-Journals}
\bibliography{bib_discrimination}





\end{document}